\documentclass[runningheads,anonymous=false]{llncs}
\usepackage{graphicx}
\usepackage{xcolor}
\usepackage{amsmath}
\usepackage{hyperref}
\usepackage{pdfpages}

\begin{document}
\title{Expert Finding in Legal Community Question Answering}
\author{Arian Askari\inst{1} \and
Suzan Verberne\inst{1} \and
Gabriella Pasi\inst{2}
}
\authorrunning{A. Askari, S. Verberne, G. Pasi}
\institute{Leiden Institute of Advanced Computer Science, Leiden University \email{first\_initial.lastname@liacs.leidenuniv.nl} \and Department of Informatics, Systems and Communication, University of Milano-Bicocca \email{gabriella.pasi@unimib.it}}
\maketitle 
\begin{abstract}
Expert finding has been well-studied in community question answering (QA) systems in various domains. However, none of these studies addresses expert finding in the legal domain, where the goal is for citizens to find lawyers based on their expertise. In the legal domain, there is a large knowledge gap between the experts and the searchers, and the content on the legal QA websites consist of a combination formal and informal communication. In this paper, we propose methods for generating query-dependent textual profiles for lawyers covering several aspects including sentiment, comments, and recency. We combine query-dependent profiles with existing expert finding methods. Our experiments are conducted on a novel dataset gathered from an online legal QA service. We discovered that taking into account different lawyer profile aspects improves the best baseline model. We make our dataset publicly available for future work.
\end{abstract}
\section{Introduction}\label{sec:intro}
Expert finding is an established problem in information retrieval \cite{balogExpertfindingBook} that has been studied in a variety of fields, including programming \cite{dehghanQuality2020,Zhou2014}, social networks \cite{WWW2020SocialExpertFinding,WWW2014SocialExpertFinding}, bibliographic networks \cite{bibNetworkExperFinding2014,bibNetworkExperFinding2018}, and organizations \cite{organizationExpertFinding2017}. Community question answering (CQA) platforms are common sources for expert finding; a key example is Stackoverflow for expert finding in the programming domain \cite{rostamiExpertFindingProgramming}.
\par
Until now, no studies have addressed expert finding in the legal domain. On legal CQA platforms, citizens search for lawyers with specific expertise to assist them legally. A lawyer's impact is the greatest when they work in their expert field \cite{SandefurElementsofProfessional}. In terms of expertise and authority, there is a large gap between the asker and the answerer in the legal domain, compared to other areas. For instance, an asker in programming CQA is someone who is a programmer at least on the junior level, and the answerer could be any unknown user. In legal CQA, the asker knows almost nothing about law, and the answerer is a lawyer who is a professional user. The content in legal CQA is a combination of formal and informal language and it may contain emotional language (e.g., in a topic about child custody). As a result, a lawyer must have sufficient emotional intelligence to explain the law clearly while also being supportive \cite{KeriEmotionalIntelligence2017}.
\par
A lawyer's expertise(s) is crucial for a citizen to be able to trust the lawyer to defend them in court \cite{lawyersPower}. Although there are some platforms in place for legal expert finding (i.e, \href{http://avvo.com/find-a-lawyer}{Avvo}, \href{http://nolo.com/}{Nolo}, and \href{http://e-justice.europa.eu}{E-Justice}), there is  currently no scientific work addressing the problem.
\par
In this paper, we define and evaluate legal expert finding methods on legal CQA data. We deliver a data set that consists of legal questions written by anonymous users, and answers written by professional lawyers. Questions are categorized in different categories (i.e, bankruptcy, child custody, etc.), and each question is tagged by one or more expertises that are relevant to the question content. Following prior work on expert finding in other domains \cite{earlydetectDavid2015,SkillTranslationSig17,dehghanQuality2020}, we select question tags as queries. We represent the lawyers by their answers' content. For a given query (required expertise), the retrieval task is to return a ranked list of lawyers that are likely to be experts on the query topic. As ground truth, we use lawyers' answers that are marked as best answers as a sign of expertise.
\par
Our contributions are three-fold: (1) We define the task of \textit{lawyer finding} and release a test collection for the task;\footnote{The data and code is available on \href{https://github.com/arian-askari/EF_in_Legal_CQA}{https://github.com/EF\_in\_Legal\_CQA}} (2) We evaluate the applicability of existing expert finding methods for lawyer finding, both probabilistic and BERT-based; (3) We create query-dependent profiles of lawyers representing different aspects and show that taking into account query-dependent expert profiles have a great impact on BERT-based retrieval
on this task. 
\section{Related work}\label{sec:rw}
The objective of expert finding is to find users who are skilled on a specific topic. The two most common ways to expert finding in CQA systems are topic-based and network-based. Because there is not a network structure between lawyers in legal CQA platforms, we focus on topic-based methods. The main idea behind topic-based models \cite{balog2009language,WSDM2020RNN,7Jinwen,20Fatemeh,13Xuebo,26Fei,28Liu} is to rank candidate experts according to the probability $p(ca|q)$, which denotes the likelihood of a candidate $ca$ being an expert on a given topic $q$. According to Balog et al. \cite{balog2009language}, expert finding can be approached by generative probabilistic modelling based on candidate models and document models. Recently, Nikzad et al. \cite{BERTERS2021} introduces a multimodal method on academic expert finding that takes into account text similarity using transformers, the author network, and h-index of the author. We approach lawyer finding differently since a lawyer does not have an h-index, there is not a sufficiently dense network of lawyers in the comment sections of legal CQA platforms, and the content style in academia is different than in legal.
\section{Data collection and preparation}\label{sec:data}
\paragraph{Data source and sample.} Our dataset has been scraped from the \href{https://www.avvo.com/topics/bankruptcy/advice}{Avvo QA forum}, which contains $5,628,689$ questions in total. In order to preserve the privacy of users, we stored pages anonymously without personal information and replaced lawyer names by a number. Avvo is a legal online platform where anyone could post their legal problem for free and receive responses from lawyers. It is also possible to read the answers to prior questions. Lawyers' profiles on Avvo have been identified with their real name, as opposed to regular users. The questions are organised in categories and each category (i.e. `bankruptcy') includes questions with different category tags  (i.e. `bankruptcy homestead exemption'). For creating our test collection, we have selected questions and their associated answers categorised as `bankruptcy' for California, which is the most populated state of the USA. We cover the period July 2016 until July 2021 which covers $9,897$ total posts and $3,741$ lawyers. The average input length of a candidate answer is $102$ words.
\par
\paragraph{Relevance labels and query selection.} We mark attorneys as experts on a \textit{category tag} when two conditions are met. The first is engagement filtering: Similar to the definition proposed in \cite{earlydetectDavid2015}, a lawyer should have ten or more of their answers marked as accepted by the asker on a \textit{category}, and a more than average number of best answers among lawyers on that \textit{category tag}. A best answer is either labelled as the most useful by the question poster or if more than three lawyers agree that the answer is useful. Second, following the idea proposed in \cite{characterisationJie2014}, the acceptance ratio (count of best answers/count of answers) of their answers should be higher than the average acceptance ratio (i.e.  $4.68\%$) in the test collection on a category. Based on the two conditions, we select $61$ lawyers as experts, who combined have given $5,614$ answers and $1,917$ best answers. From the top $20$ percent tags which co-occur with `bankruptcy', we select tags ($84$) as queries that at least have two experts. There are on average $5$ experts (lawyers who met expert conditions on a \textit{category tag}) per query in the test collection. Our data size is comparable with four TREC Expert Finding test collections between 2005-2008, that have $49$--$77$ queries and $1,092$--$3,000$ candidates \cite{craswell2005overview,soboroff2006overview,bailey2007overview,balog2008overview}.

\paragraph{Evaluation setup.} 
We split our data into train, validation, and test sets based on the relevant expert lawyers -- instead of queries -- to avoid our models being overfitted on previously seen experts. By splitting on experts, the retrieval models are expected to be more generalized and be able to detect new experts in the system. The distribution of relevant experts and queries in each set is shown in table \ref{tab:stat_sets}. For each train/valid/test set, in retrieval, we have all non-relevant lawyers ($3680$ in total) plus relevant lawyers (experts) ($20/20/21$) to be ranked.
\begin{table}[h]
\centering
\caption{Statistics on the counts of queries, answers, and relevant experts in our data.}\label{tab:stat_sets}
{\scriptsize
\begin{tabular}{l|l|l|l|l|l}
        & train & validation & test & train $\cap$ validation  & train $\cap$ test \\ \hline
number of relevant experts & 20    & 20  & 21   & 0           & 0            \\ 
number of queries     & 76    & 69  & 71   & 61          & 65           \\ 
number of answers     &  39,588  &  34,128 &  35,057  &         7,290  &       7,918 \\
\hline
\end{tabular}
}
\end{table}
\section{Methods}\label{sec:method}
Lawyer finding is defined as finding the right legal professional lawyer(s) with the appropriate skills and knowledge within a state/city. Cities are provided by Avvo as metadata; we only keep the city of the asker in our ranking and filter out lawyers' answers from other cities. For relevance ranking, lawyers are represented by their answers, like in prior work on expert finding in other domains \cite{balog2009language}.
\subsection{Baseline 1: Probabilistic language modelling}
Following \cite{earlydetectDavid2015,SkillTranslationSig17,dehghanQuality2020}, we replicate two types of probabilistic language models to rank lawyers: document-level (model $1$), and candidate-level (model $2$) that were originally proposed by Balog et al. \cite{balog2009language} In these models, the set of answers written by a lawyer is considered the proof of expertise. 
\par
In the \textbf{Candidate-based model,} we create a textual representation of a lawyer's knowledge based on the answers written by them. Following Balog et al. \cite{balog2009language}, we estimate $p(ca|q)$ by computing $p(q|ca)$ based on Bayes’ Theorem. We call this model hereinafter \textit{model 1}. In \textit{model 1}, $P(q|ca)$ is estimated by:
\begin{equation}
    p(q|ca)=\prod_{t\in{q}}{\bigg\{ (1-\lambda_{ca}) \times \bigg( \sum_{d\in D_{ca}}{p(t|d) \times p(d|ca)} \bigg) + \lambda_{ca} \times p(t) \bigg\}}
\end{equation}
\par
Here, $D_{ca}$ consists of documents (answers) that have been written by lawyer $ca$; $p(t|d)$ is the probability of the term $t$ in document $d$; $p(t)$ is the probability of a term in the collection of documents; and $p(d|ca)$ is the probability of document $d$ is written by candidate $ca$. In the legal CQA platform answers are written by one lawyer. Therefore, $p(d|ca)$ is constant.
\par
In the \textbf{Document-based model}, the document-centric model builds a bridge between a query and lawyers by considering documents in the collection as link. Given a query $q$, and collection of answers ranked according to $q$, lawyers are ranked by aggregating the sum over the relevance scores of their retrieved answers:
\begin{equation}
    p(q|ca)= \sum_{d\in D_{ca}} \Bigg( \prod_{t\in{q}}\bigg\{ (1-\lambda_d)\times p(t|d) + \lambda_d \times p(t) \bigg\} \times p(d|ca) \Bigg)
\end{equation}
\par
$\lambda_{d}$ and $\lambda_{ca}$ are smoothing parameters that are dynamically computed per query and candidate lawyer document (lawyer's answer)/representation following \cite{balog2009language}. Besides of the original \textit{model 1}, and \textit{model 2} based on probabilistic language modelling, we experiment with BM25 to rank expert candidates' profiles and documents and refer to those by \textit{model 1 BM25}, and \textit{model 2 BM25}.

\subsection{Baseline 2: Vanilla BERT}
By Vanilla BERT, we mean a pre-trained BERT model (BERT-Base, Uncased) with a linear combination layer stacked atop the classifier [CLS] token that is fine-tuned on our dataset in a pairwise cross-entropy loss setting using the Adam optimizer. We used the implementation of MacAvaney et al. \cite{CEDR2019} (CEDR).
\par
After initial ranking with \textit{model 2}, we fine-tune Vanilla BERT to estimate the relevance between query and answer terms. We select retrieved answers of the top-k($50$) lawyers to re-calculate their relevance score by Vanilla BERT according to the query $q$. Finally, we re-rank the top-k by these relevance scores. Given a query and an answer, we train Vanilla BERT to estimate the relevance that the answer was written by an expert: ``$ \texttt{[CLS]query[SEP]candidate answer[SEP]}$".
\subsection{Proposed Method}
Given a query $q$ and a collection of answers $D$ that are written by different lawyers, we retrieve a ranked list of answers ($D_q$) using \emph{model 1}. We create four query-dependent profiles for the lawyers $L_q$ who have at least one answer in $D_q$. Each profile consists of text, and that text is sampled to represent different aspects of a lawyer's answers. The aspects are comments, sentiment-positive, sentiment-negative, and recency.
\par
On the CQA platform it is possible to post comments in response to lawyer's answer. Therefore, there is a collection of comments $C_{D_q}$ with regard to the query. We consider the comments as possible signals for the asker's satisfaction (i.e., a ``thank you'' comment would indicate that the asker received a good answer). Thus, for \textbf{comment-based profiles ($CP$)}, we shuffle the comments to $l_i$'s answers and concatenate the first sentence of each comment. For \textbf{sentiment-positive ($PP$) and negative ($NP$) profiles}, we shuffle positive (negative) sentences from $l_i$'s answers and concatenate them. Since our data in legal CQA is similar in genre to social media text, we identify answer sentiment using Vader \cite{vaderSentiment}, a rule-based sentiment model for social media text. For the \textbf{recency-based profile ($RP$)}, we concatenate the most recent answers of $l_i$. For each profile we sample the text until it exceeds $512$ tokens.
\par
We fine-tune Vanilla BERT on each profile. We represent the query as sentence A and the lawyer profile as sentence B in the BERT input:
   `` \texttt{[CLS] \enspace query \enspace [SEP] \enspace 
    lawyer \enspace profile  \enspace [SEP]}''
\begin{table}[t]
\caption{Baselines and proposed model results on the test set. Significant improvements over the probabilistic baselines (Model 1 LM/Bm25, Model 2 LM/BM25), and over the Vanilla BERT Document-based models are marked with \textasteriskcentered, and \textbullet respectively.}\label{tab:ret_res} .
\centering
{\scriptsize
\begin{tabular}{l|c|c|c|c|c}
Model       & MAP                         & MRR                         & P@1                         & P@2                         & P@5                   
\\ \hline\hline
Model 1 (Candidate-based) LM                         & 22.8\%                     & 40.9\%                     & 23.9\%                     & 19.7\%                     & 13.2\%                      
\\ \hline
Model 1 (Candidate-based) BM25                      & 3.7\%                     & 7.0\%                     & 2.9\%                     & 1.4\%                     & 2.6\%                     
\\ \hline
Model 2 (Document-based) LM                          & 19.4\%                     & 21.9\%                     & 13.5\%                     & 12.7\%                     & 7.8\%                     
\\ \hline
Model 2 (Document-based) BM25 & \multicolumn{1}{c|}{21.0\%} & \multicolumn{1}{c|}{36.6\%} & \multicolumn{1}{c|}{22.5\%} & \multicolumn{1}{c|}{18.3\%} & \multicolumn{1}{c}{11.5\%}  
\\ \hline
Vanilla BERT Document-based (VBD) & \multicolumn{1}{c|}{37.3\%*} & \multicolumn{1}{c|}{70.7\%\textasteriskcentered} & \multicolumn{1}{c|}{60.5\%\textasteriskcentered} & \multicolumn{1}{c|}{55.6\%\textasteriskcentered} & \multicolumn{1}{c}{25.9\%\textasteriskcentered} 
\\ \hline
VBD + Profiles  (weighted) & \multicolumn{1}{c|}{\textbf{39.3}\%\textasteriskcentered\textbullet} & \multicolumn{1}{c|}{\textbf{73.2}\%} & \multicolumn{1}{c|}{\textbf{64.9}\%\textasteriskcentered\textbullet} & \multicolumn{1}{c|}{\textbf{57.1}\%} & \multicolumn{1}{c}{\textbf{27.7}\%\textasteriskcentered\textbullet}
\\ \hline
\end{tabular}
}
\end{table}
Finally, we aggregate the scores of the four profile-trained BERT models and \textit{BERT Document-based} using a linear combination of the five models' scores inspired by \cite{althammer2021dossier}: $aggr_{S}(d,q) = w_1 S_{BD} + w_2 S_{CP} + w_3 S_{PP} + w_4 S_{NP} + w_5 {RP}$, where $aggr_{S}$ is the final aggregated score; the weights $w_i$ are optimized using grid search in the range $[1..100]$ on the validation set. $BD$ refers to the \textit{BERT Document-based} score, and $CP$, $PP$, $NP$, $RP$ to the four profile-trained BERT models. 
\section{Experiments and Results}
\paragraph{Experimental setup}
We replicate \cite{balog2009language} using Elasticsearch for term statistics, indexing, and BM25 ranking.
Following the prior work on expert finding, we report MAP, MRR, and Precision@k ($k=1,2,5$) as evaluation metrics.
\paragraph{Retrieval results}
The ranking results for models are shown in table \ref{tab:ret_res}. The best candidate-based and document-based lexical models are the original \textit{model 1 LM} \cite{balog2009language}, and \textit{model 2 BM25} respectively. We used \textit{model 2 BM25} as our initial ranker for Vanilla BERT. \textit{Vanilla BERT Document-based}  outperforms all lexical models by a large margin. The best ranker in terms of all evaluation metrics is the weighted combination of BERT and the lawyer profiles. This indicates that considering different aspects of a lawyer's profile (comments, sentiment, recency) is useful for legal expert ranking. We employed a one-tailed t-test ($\alpha = 0.05$) to measure statistical significance.
\paragraph{Analysis of models' weights.} We found $20,13,2,4,1$ as optimal weights for BERT, Comment, Recency, Sentiment positive and negative based models respectively. As expected, the BERT score plays the largest role in the aggregation as it considers all retrieved answers of a lawyer. The second weight is for the Comment profile which confirms our assumption that the content of askers' comments are possible signals for the relevance of the lawyer's answer. The Sentiment profile's weight
shows positive sentiment is more informative than negative on this task.
\paragraph{Analysis of differences on seen and unseen queries.} In Section~\ref{sec:data}, we argued that in our task, being robust to new lawyers is more important than being robust to new expertises (queries). We therefore split our data on the expert level and as a result there are overlapping queries between train and test set. We analyzed the differences in model effectiveness between seen and unseen queries. We found small differences: $p@5$ is $27\%$ on seen queries, and $25\%$ on unseen queries. This indicates the model generalizes quite well to unseen queries.
\section{Conclusions}
In this paper, we defined the task of legal expert finding. We experimented with baseline probabilistic, BERT-based, and proposed expert profiling methods on our novel data. BERT-based method outperformed probabilistic methods, and the proposed methods outperformed all models.
\par
For future work, there is a need to study more in-depth the robustness of proposed methods on different legal categories. Moreover, by providing this dataset we facilitate other tasks such as legal question answering, duplicate question detection, and finding lawyers who will reply to a question. 
\section*{ACKNOWLEDGMENTS}
This work was supported by the EU Horizon 2020 ITN/ETN on Domain Specific Systems for Information Extraction and Retrieval (H2020-EU.1.3.1., ID: 860721).
\bibliographystyle{splncs04}
\bibliography{ref.bib}
\end{document}